\documentclass[aps,pre,twocolumn,groupedaddress]{revtex4}
\usepackage{amsmath}
\usepackage{graphicx}
\newcommand{\fref}[1]{Fig.~\ref{#1}}

\newcommand{\eref}[1]{(\ref{#1})}
\newcommand{\degree}{\ensuremath{^\circ}}

\begin{document}

\title{A direct optical method for the study of grain boundary melting}


\author{E. S. Thomson}
\email[]{erik.thomson@yale.edu}
\affiliation{Department of Geology and Geophysics, Yale University, New Haven, CT, 06520, USA}

\author{J. S. Wettlaufer}
\affiliation{Departments of Geology \& Geophysics and Physics, Program in Applied Mathematics, Yale University, New Haven, CT, 06520, USA}
\date{\today}

\author{L. A. Wilen}
\affiliation{Department of Geology and Geophysics, Yale University, New Haven, CT, 06520, USA}
\affiliation{Unilever Research Corporation, Trumbull, CT, 06611, USA}
\date{\today}

\begin{abstract}
The structure and evolution of grain boundaries underlies the nature of polycrystalline materials.  Here we describe an experimental apparatus and light reflection technique for measuring disorder at grain boundaries in optically clear material, in thermodynamic equilibrium.  The approach is demonstrated on ice bicrystals.  Crystallographic orientation is measured for each ice sample.   The type and concentration of impurity in the liquid can be controlled and the temperature can be continuously recorded and controlled over a range near the melting point.  The general methodology is appropriate for a wide variety of materials.

\emph{Copyright 2009 American Institute of Physics. This article may be downloaded for personal use only. Any other use requires prior permission of the author and the American Institute of Physics.  The following article appeared in Review of Scientific Instruments (Vol.80, Issue 10) and may be found at} \url{http://link.aip.org/link/?RSI/80/103903}.
\end{abstract}


\maketitle
\section{Introduction}

The phase behavior and geometry of polycrystalline materials is of fundamental scientific interest.  Within polycrystals, the structure, number, and evolution of grain boundaries control bulk mechanical and electrical properties \cite{Dai2009}.  Grain boundaries are also accountable for grain growth, recrystallization, and other phenomena underlying the physical character of polycrystalline materials \cite{Konijnenberg2008}.  Because polycrystalline materials are ubiquitous in engineering applications, like liquid crystals \cite{Kleman2008} and ceramics \cite{Luo2008}, and in geophysical contexts, such as the partially molten inner earth \cite{Frank1968}, grain boundaries remain an important area of research.   

In general, the accessibility of microscopic grain boundaries, which are internal to the polycrystalline phase geometry, challenges experimental studies.  However, the transparency and birefringence of ice make it amenable to laboratory investigation.   Furthermore, ice is of particular geophysical interest because it is commonly found at or near phase coexistence.  Previous studies of ice's phase geometry have largely focused on free surfaces \cite{Elbaum1993, Engemann2004} and the macroscopic liquid network of veins and nodes \cite{Nye1973}.  These veins and nodes are large ($10-100\; \mu $m) conduits of liquid water threading throughout the solid polycrystal where three or more grains intersect and can be directly observed with optical microscopy \cite{Mader1992}.  Measurements of surface energy \cite{Ketcham1969} and grain boundary grooves \cite{Walford1991} have also contributed to the understanding of ice's phase behavior.  More recently, focus has turned to the interface between two single grains, what we define as a grain boundary.  At grain boundaries experiments have focused on mobility, diffusion, and the dihedral angle (for a complete review of the literature in ice see \cite{Dash2006}).   Examples include a study by Nasello et al. \cite{Nasello2007} of grain boundary migration rates.  Their observations suggest that grain boundary structure depends upon the annealing temperatures and solute concentration.  Another study \cite{Lu2007} used an ablation technique to make conclusions concerning rates of diffusion and therefore disorder at grain boundaries.  In this technique layered samples with regions of isotopically heavy water are ablated.  The ablated material is analyzed using a mass spectrometer and the diffusive profile around the isotopically heavy regions is correlated with the annealing time.  Such measurements provide a wide range of valuable evidence from which aspects of the character and the scale of grain boundary disorder can be inferred. The goal of our complimentary approach is to probe the grain boundary directly.

Theory \cite{Benatov2004} shows that as the strength of intermolecular interactions fluctuates, melting at grain boundaries in ice may be discontinuous, and highly sensitive to ionic impurities. While it is commonly assumed that the observation of a finite dihedral angle is necessary and sufficient to rule out melting at a grain boundary \cite{Caupin2008}, the disparity of scales between capillary lengths (observable with optical microscopy) and interfacial disorder across a grain boundary can be enormous.  The full description of this range of scales requires keeping the interaction term in the free energy.  A disordered layer thickness $\ll1$ wavelength cannot be measured using conventional optical methods.  Hence a direct reflectance technique has been developed.  

We have designed an experimental ice growth cell and light reflection technique that provides direct optical access to grain boundaries.  Ice crystal orientations and reflected light intensity are measured in thermodynamic equilibrium.  The complicated behavior of plane wave propagation through an isotropic layer between uniaxial crystals has been theoretically modeled \cite{Thomson2009} and the observed behavior, taking into account the measured crystallography, corresponds well with the theoretical analysis.

\section{Experimental Design}

In opaque materials direct access to grain boundaries is largely obstructed.  Conversely, bubble free pure ice crystals are transparent and relatively easy to control near the triple point.  Our design allows continuous monitoring of the intensity of light reflection from a grain boundary as thermodynamic variables are changed.  The approach incorporates temperature measurement and control, impurity control, a method for crystallographic orientation measurement, and optical access for light.  Here we describe the elements of the experimental apparatus, and a procedure by which it may be used in order to characterize quantitatively the thermodynamically stable grain boundary structure in ice.  The design is sufficiently general that it can translate to a range of other optically clear materials.    

\subsection{Ice Growth Apparatus}

\begin{figure}\centering
\includegraphics[scale=1.0]{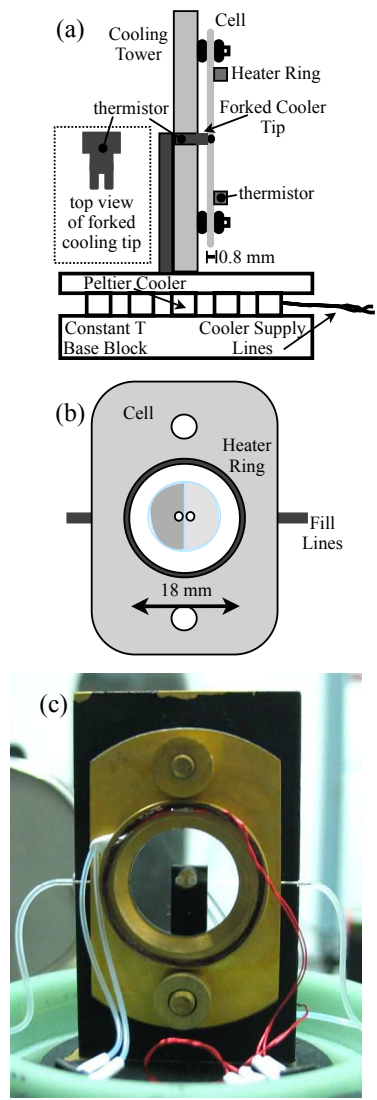}
\caption{(a) A side view schematic of the ice growth cell mounted on the cooling tower, with a view of the cooler tip from above, inset.  (b) A schematic of the ice growth cell with an example ice disk drawn inside the sealed disk-like chamber.  (c) A picture of the mounted cell with attached flow lines.  Color available online.}
\label{fig:cell}
\end{figure}

The core mechanical components are the ice growth cell and a cooling stage mount, generalized from the apparatus of Wilen and Dash \cite{Wilen1995a, Wilen1995b}.  Our design, shown in \fref{fig:cell}, is oriented vertically, and incorporated into a coherent optical beam line.  

The ice growth cell is formed using a 0.8 mm thick wafer through which an 18 mm hole is bored.  Small holes are drilled through each side of the wafer into the thin disk-like region, into which $\approx 0.6$ mm diameter flow tubes are inserted. The wafer is sandwiched by two glass cover slips and sealed with paraffin, creating an empty cylindrical volume.  A resistance heating ring, encircling the outside diameter of the bored hole is mounted to the exterior of the wafer/cover slip system. Water is flowed into the sealed disk-like chamber through fill lines connected to the flow tubes, creating a liquid reservoir.  The entire cell is mounted onto a cooling unit and ice nucleated within the reservoir.  

The cell with the attached heater ring is mounted vertically onto a machined stand cooled from below using a thermoelectric cooling device.  Thermal grease mechanically connects a cooling finger from the stand to the center of one of the glass cover slips.  This central cell is enclosed within an insulating case with optical windows to allow for the laser beam propagation.  Dry nitrogen is continuously pumped through lines heat sunk to the cooling tower into the insulating shell for thermal buffering and desiccation.  The cell fill lines thread through the insulating shell to connect the cell with external reservoirs of solution.  Similarly the thermistor leads, and cooler and heater power lines connect with the temperature control assemblage.  

For the grain boundary experiments, a forked cooling finger has been engineered to aid in the formation of  an oriented bicrystal.  This split cooler tip provides two individual points of contact with the glass cover slip.  By adjusting the temperature these two points allow isolation of two single ice grains with different crystallographic orientations.  This procedure yields the desired vertically oriented grain boundary.

\begin{figure}\centering
\includegraphics[scale=1.0]{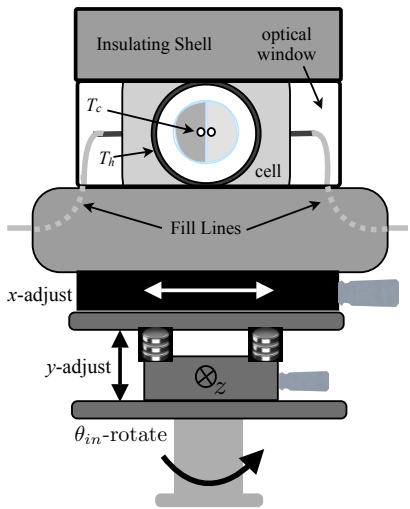}
\caption{Schematic of the ice growth cell enclosed in the insulating shell, mounted upon translation and rotation stages.  These adjustments allow the focus point and incident angle to be adjusted. }
\label{fig:adjs}
\end{figure}

This entire system is mounted on rotation and translation stages (\fref{fig:adjs}) in a laser beam's path.  The horizontal translation stage allows for fine lateral adjustment of the beam's path through the grain boundary.  The vertical adjustment enables the experimenter to probe different grain boundary temperatures.  Finally, the rotational degree of freedom allows for adjustment of the beam incident angle and is also important for the identification of crystallographic orientation (Sec. \ref{sec:xgraphy}).

\subsection{Temperature Measurement and Control}

The temperature of the cooling unit is regulated using a Conductus$^{\tiny\textregistered}$ LTC-10 temperature control module in conjunction with a thermoelectric cooling device.  One k$\Omega$ YSI precision thermistors, embedded into the metal of the cooler tip and heater ring, monitor the temperatures at the center and outside diameter of the growth cell.  The resistance heater ring is powered directly by the Conductus$^{\tiny\textregistered}$ DC heater output.  The low current analog output loop of the Conductus$^{\tiny\textregistered}$ is coupled to a higher current power supply that drives the thermoelectric device cooling the base unit.  The warm side of this peltier device is kept at a constant temperature using a NesLab$^{\tiny\textregistered}$ recirculating bath.  

The central cold finger and exterior heater ring, combined with the aspect ratio of the cell ($\approx 0.04$), create, to a good approximation, a two-dimensional radial temperature gradient.  Temperatures can be measured to an accuracy limited by the thermistors, and are stably controlled to a few millidegrees.  Simultaneously with light scattering data, the set point and measured temperatures are downloaded from the Conductus$^{\tiny\textregistered}$ using LabView$^{\tiny\textregistered}$.  The temperature at the beam measurement location is calculated using a temperature model for the ice cell described in the Appendix.  Using the model, temperature at the beam measurement location is calculated, at the $95\%$ confidence levels, to within $0.02 K$. 

\subsection{Nucleation}

\begin{figure}
\centering
\includegraphics[scale=1.0]{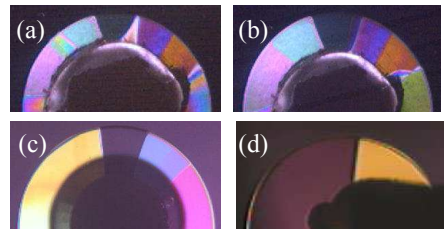}
\caption{Images of the upper half of ice disks, with central regions obscured by the cooler tip, viewed between light polarizing sheets representing the stages of nucleation and coarsening as an ice disk is annealed to isolate a bicrystal.  (a) Example of a recently nucleated ice disk, where the many orientations are clear.  (b) After short annealing times some orientations are removed, grain boundaries become straighter and more abrupt as the interfacial energy is minimized.  (c) Example of a largely annealed ice disk with a handful of crystal orientations remaining.  (d) An example sample coarsened until only two orientations with a single vertical grain boundary are present.  Each of these stages may be accelerated using freeze/thaw cycles, however, the fundamental process remains the same.}
\label{fig:gbimage}
\end{figure}

In all experiments, pure degassed water is initially flowed into the disk-like sample chamber.  Ice is nucleated from this pure water by pressing a cotton swab cooled in liquid nitrogen against the cover slip opposite the cooler tip in order to strongly supercool the liquid.  A distribution of orientations are nucleated in any single event.   Subsequent temperature cycling results in an ice disk of two or more crystal domains with radial grain boundaries separating individual crystals like spokes on a bicycle wheel (\fref{fig:gbimage}).  For the grain boundary light scattering experiments, two of these orientations are isolated by warming and cooling cycles which anneal the crystals; ultimately individual seed crystals remain at each tip of the bifurcated cooler finger and are re-grown into a bicrystal, with an optically accessible grain boundary.   Once formed the orientation of the crystals is measured using a method discussed in Sec. \ref{sec:xgraphy}.

\subsection{Impurity Concentration}

Impurities can be introduced into the liquid using the flow tubes.  However, because nucleation in an impurity rich solution introduces interfacial instabilities \cite{Dash2006}, the annealing of which is a very slow process, we generally choose to nucleate with a pure reservoir.  Subsequent to nucleation the concentration and type of impurity within the annular region of bulk liquid surrounding the ice may be changed by flowing a new solution through the fill lines.  Any type of impurity which will not block the fill lines can be used.  Indeed this type of ice growth cell is suited to the study of many impurity effects including experiments in cryobiology \cite{Pertaya2007a}.  However, in order to compare measurements of ice grain boundaries with theoretical calculations \cite{Benatov2004} simple monovalent electrolytes are used.  

The flow of impurity into the cell is done carefully in an effort to minimize the perturbation to the ice/water interface and temperature control is adjusted to maintain a constant temperature profile, relative to the melting temperature of the bulk solution.  Unfortunately, at the scale of the cell it is impossible to replace the liquid while keeping the entire system at equilibrium.  The Peclet number, or ratio of advection to diffusion within a system, is 
\begin{equation}
\label{eq:peclet} \text{Pe}=\frac{LV}{\alpha},
\end{equation}
 where $L$ is the characteristic length scale, $V$ the velocity, and $\alpha$ is a diffusivity.  For the experimental cell, $L\sim10^{-3} m$ and $\alpha \sim 10^{-7}-10^{-9}\; m^2/s$, depending upon whether the diffusive property is heat or solute.  In order to have a Peclet number smaller than one requires unrealistically slow flow rates.  Therefore, the solution is flowed into the cell in pulses which are sufficiently slow to limit the thermal input to the system, but fast relative to solutal diffusion.  Each flow pulse completely replaces the volume of the cell and attached tubing at least five times.  The interface initially melts back, due to the influx of solute, and eventually regrows slowly to its equilibrium position.  As this occurs, the cooler tip and heater ring temperatures are adjusted to keep them constant relative to the melting temperature.  The entire system re-equilibrates for over one hour, the time scale for solutal diffusion.  Flow is controlled using a gravity feed system, or more precisely using a chain of syringe pumps. 

\subsection{Crystallographic Measurements}
\label{sec:xgraphy}

\begin{figure}
\centering
\includegraphics[scale=1.0]{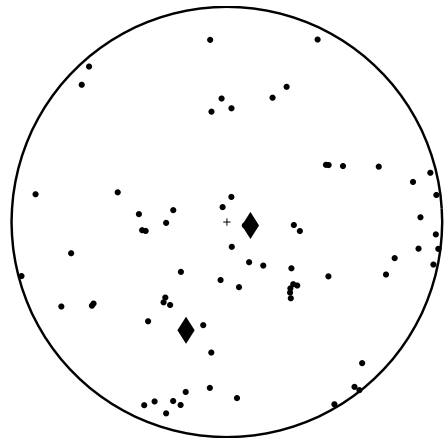}
\caption{Equal area projection (often called a Schmidt plot) of unit vectors representing c-axis orientations of crystals nucleated and measured within the ice growth cell.  Clearly, there are no preferred orientations nucleated within the cell.  Taken cumulatively points could represent a recently nucleated, highly polycrystalline sample, with the two diamonds indicating an annealed bicrystal.} 
\label{fig:sch}
\end{figure}

By definition grain boundaries separate misoriented crystal domains.  Our experimental apparatus does not allow us to \emph{a priori} choose a particular mismatch of orientations.  Rather, the process of nucleation generates a random distribution of orientations (e.g., \fref{fig:sch}).  From that random distribution the annealing process isolates a bicrystal, represented by the pair of diamonds in the Schmidt plot (\fref{fig:sch}).  The magnitude of the crystallographic mismatch across a grain boundary is subsequently determined by independently measuring each crystal's orientation.  The measurements are made using a technique modeled upon one developed for analyzing the ice fabrics of large ice core sections \cite{Wilen2000} and adapted to the geometry and constraints of our experimental ice growth cell in order to allow all orientation measurements to be made \emph{in situ}. 

The principle of the technique \cite{Wilen2000}, informed by earlier treatments of universal stage measurements \cite{Kamb1962}, utilizes the uniaxial nature of the crystal.  As a plane polarized wave traverses an ice crystal its polarization is rotated. If the wave subsequently passes through a polarizing element oriented perpendicular to the incident wave polarization, the intensity of the transmitted light will be related to the mismatch of the crystal's optic axis and the initial and final polarizations.  Clearly, if there is no rotation then no light is transmitted through the system because the two polarizations are orthogonal.  This is only true when the optical axis of the intervening crystal is aligned with, or perpendicular to the plane determined by one of these polarizations.  By varying the orientation of the crystal sample within the lab reference frame, and using extinction angles to determine multiple intersecting planes, which may include the c-axis, the direction of the axis can be uniquely determined.  

\begin{figure}
\centering
\includegraphics[scale=1.]{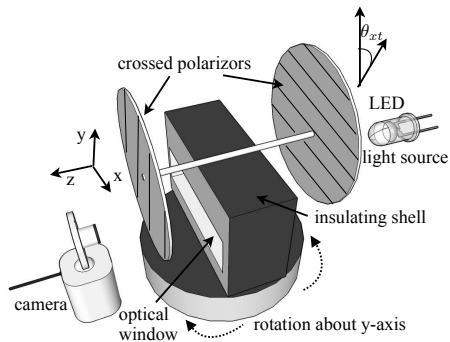}
\caption{Schematic of the laboratory assembly for measuring c-axis orientations.  A LED array transmits white light through the insulating cover's optical windows and enclosed ice growth cell to a camera.  The cell is straddled by permanently crossed polarizing sheets that are rotated simultaneously and the entire cell is free to rotate about the y-axis. Optional wedges can be used to add a rotation about the z-axis.  Extinction angles ($\theta_{xt}$) are recorded for each crystal and cell position, and a minimization scheme \cite{Wilen2000} is used to solve for c-axis orientation.  } 
\label{fig:caxmeas}
\end{figure}

To apply this technique we insert a white light source, digital camera, and large rotating crossed polarizing sheets, straddling the ice growth cell (\fref{fig:caxmeas}).  The horizontal rotation stage allows the orientation of the crystal, within the laboratory frame to be changed, and further tilt can be added by using machined $45\degree$ wedges.  The digital camera is interfaced with a computer and transmitted light intensity from a selected region is monitored using LabView$^{\tiny\textregistered}$ image tools.  This is done at fixed cell positions as the crossed polarizers are simultaneously rotated and their position recorded at the transmitted light intensity minimum for single grains.  The procedure is repeated for multiple crystal positions enabling us to solve for c-axis position using the minimization scheme outlined in \cite{Wilen2000}.   

\subsection{Optics}

\begin{figure}
\centering
\includegraphics[scale=1.0]{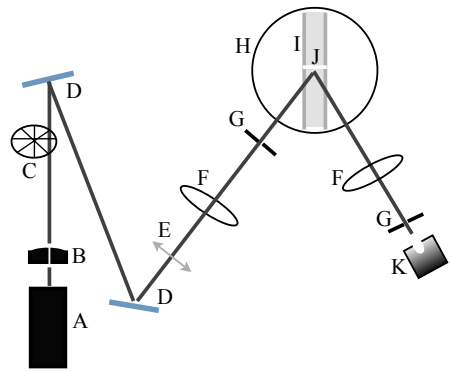}
\caption{Schematic of the optical bench set up.  The laser beam is spatial filtered and frequency chopped before the polarization is set and the beam is focused onto the grain boundary.  The reflected signal is refocused onto a calibrated photo detector. A - laser, B - spatial filter, C - optical chopper D - mirrors, E - polarizor, F - lenses, G - pinholes, H - environmental chamber, I - cell, J - grain boundary, K - photo detector.}
\label{fig:bench}
\end{figure}

Finally we return to the light reflection portion of the experimental system.  In order to make reflection measurements the thermally insulated ice growth apparatus, mounted onto the rotation and translation stages, is placed in the path of a laser beam on an optical bench.  With the cell in this position, the beam is focused onto the grain boundary and the reflected signal collected and analyzed.  \fref{fig:bench} illustrates the optical elements and the path of the laser beams propagation.

To limit absorption and scattering within the ice, for these experiments a 2.3 mW, 632.8 nm He-Ne laser was chosen.    Based upon an adsorption coefficient calculated using data from \cite{Warren2008} less than $0.02\%$ of the laser's beam is attenuated within the ice itself.  Thus, any heating effect is insignificant and below the resolution of our thermometry.  The light is spatially filtered and recollimated to a beam diameter of 3 mm.  From geometric optics we can calculate the beam waist diameter ($w$) for a Gaussian beam focused by a lens with focal length $F$ as,
\begin{equation}
\label{eq:bmwaist} w=\frac{2\lambda F}{\pi D},
\end{equation}
where $D$ is the diameter of the beam input into the lens and $\lambda$ is the wavelength of the light.  Using a lens with a 200 mm focal length to refocus the beam yields a beam waist diameter of approximately 30 microns.  Importantly, this beam waist diameter is an order of magnitude smaller than the thickness of the wafer (0.8 mm) and therefore the width of the grain boundary between the coverslips.  The reflected and transmitted light propagating away from the grain boundary leaves the insulating shell through optical windows.  

The incident beam is spatially filtered, frequency chopped, and polarized before it is focused onto the grain boundary.  Thus the reflected signal can be frequency-locked to the incident beam.  The reflection is refocused onto a calibrated photo detector whose output goes to a lock-in amplifier.  The frequency-locked signal and temperature data are read by a computer at specified time intervals.  Typically the time constant of the lock-in amplifier is set to be 10 seconds and the data is sampled and recorded every 60 seconds.  

\begin{figure}
\centering
\includegraphics[scale=1.0]{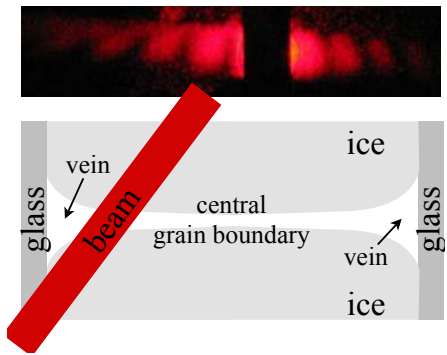}
\caption{Schematic of the grain boundary with the veins and central grain boundary indicated.  When the laser beam traverses the curved portion of the vein it acts as an unstopped lens creating a diffraction pattern in the far field.  In order to visualize the full diffraction pattern the high intensity central maxima is blocked.}
\label{fig:glvein}
\end{figure}

During initial tests of the apparatus we discovered a unique method of focusing the beam on the central grain boundary (\fref{fig:glvein}), midway between coverslips.  A schematic of the grain boundary geometry from above (\fref{fig:glvein}) shows that at the glass/ice interface there is some curvature associated with the grain boundary.   This curvature allows for the formation of a vein of liquid water where the grain boundary meets the glass.  Coincidentally, this curved region has a unique scattering signal when traversed by the laser beam.  The signal from the vein generates a strong diffraction pattern with decaying intensity away from a central maximum.   Nye \cite{Nye1991} observed this same phenomena and pointed out that within polycrystalline ice under pressure small water lenses form on veins.  When monochromatic light is incident upon one of these liquid inclusions it acts as a small unstopped lens, generating a diffraction pattern which is the superposition of the light that passes through the lens and that which passes unobstructed around the lens.   In our experimental set up when the laser light passes through the veins at an angle they act as small unstopped lenses.  The observed diffraction patterns are directly analogous to Nye's observations and provide a convenient way to locate the central grain boundary.  As the incident beam is systematically swept across the grain boundary the vein signals can be used to isolate a signal from the central grain boundary.  The geometry of the experimental setup dictates that signals of this type will straddle any signal from the inner grain boundary.     

Once the coherent beam is correctly focused, reflection intensity data is continuously collected as a function of the controlling thermodynamic variables.  For example, as the ice disk is grown or shrunk by changing the cooler tip temperature the beams position is measuring at a colder or warmer temperature.  Likewise the impurity concentration can be changed, or new bicrystals grown.  Experience shows that the most stable measurements are made by setting a beam position and subsequently varying impurity and/or temperature while continually collecting intensity data.  Physical interpretation of the reflected intensity data then requires a detailed theoretical model of reflection from the grain boundary \cite{Thomson2009}.  

\section{Demonstration of Method}

\begin{figure}
\centering
\includegraphics[scale=1.0]{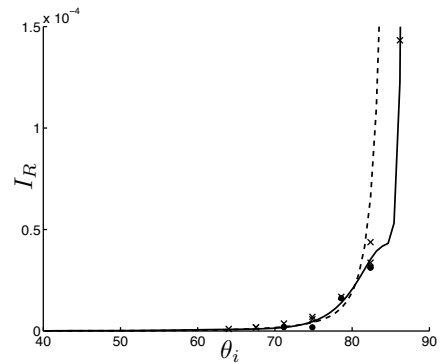}
\caption{Reflected intensity ratio $(I_R)$ versus grain boundary incidence angle $(\theta_i)$.  X's are data recorded at multiple incident angles.  The solid curve is the theoretical reflection as a function of incidence angle for the same grain boundary.  The c-axis orientations were measured and can be expressed as angles from the x,y, and z directions in the laboratory frame, $(\theta_\alpha^1,\theta_\beta^1,\theta_\gamma^1)=(52.2\degree,108.2\degree,43.46\degree)$ and $(\theta_\alpha^2,\theta_\beta^2,\theta_\gamma^2)=(10.2\degree,90\degree,79.8\degree)$. The dashed line and solid points correspond with the theory for and data from a grain boundary separating crystals with orientations given by, $(\theta_\alpha^1,\theta_\beta^1,\theta_\gamma^1)=(112.52\degree,118.07\degree,37.35\degree)$ and $(\theta_\alpha^2,\theta_\beta^2,\theta_\gamma^2)=(64.58\degree,91.09\degree,25.45\degree)$.}
\label{fig:thrywangs2}
\end{figure}

Reflected light intensity from a grain boundary depends upon crystallographic characteristics and the character of any intervening isotropic layer.  The ratio of reflected to incident intensity, $I_R$, will vary as a function of the dielectric constants of the materials, their orientations, the thickness of any liquid layer, and the incident angle of the light.  \fref{fig:thrywangs2} illustrates that measurements of the angular dependence of reflected intensity ratio show good agreement with curves calculated using theory \cite{Thomson2009}.   In this case we have nucleated a bicrystal from pure, degassed water, and added no impurities into the system.  As a result we expect immeasurable melting using the light scattering technique, and therefore use the limit of zero grain boundary thickness for the theoretical analysis.  

In order to compare the theoretical $I_R$ \cite{Thomson2009} with the experimentally measured $I_R$ the beam intensity immediately before and after the grain boundary must be known.  The incident beam intensity and attenuation due to each optical element of the system were measured using a Fieldmaster$^{\tiny\textregistered}$ power meter.  Just before propagating into the insulating shell the measured incident intensity is a constant.  Further attenuation due to the air/glass interfaces of the shell and cell (all of which depend upon angle of incidence) are taken into account using the isotropic Fresnel equations.  At the glass/ice interface attenuation is calculated using the full treatment of the isotropic/anisotropic interface \cite{Thomson2009}.  Additionally, we account for the beam's exit from the cell and insulating shell to find the reflected intensity from the measured voltage at the lock-in.  The ratio of the reflected intensity to the incident intensity can then be compared with the theoretically calculated values (\fref{fig:thrywangs2}).

The well predicted measurements of angular dependence clearly demonstrate the capabilities of this experimental apparatus used in conjunction with a complete optical theory.  We are utilizing this experimental technique in a long term effort to quantify the character of disorder at grain boundaries as a function of the fundamental thermodynamic variables; temperature, impurity content, and crystallographic orientation. 

\section{Discussion}

The experimental design we have described for measuring grain boundaries within laboratory grown crystal samples uses multiple techniques and a variety of instrumentation.  However, taken independently each measurement is straightforward and theoretically well-described.  Taken together the design enables the experimenter to characterize completely a grain boundary in ice or other materials as a function of its thermodynamic state.  

This instrumentation is also amenable to other uses, which have not been discussed but warrant attention in the scientific community.  This apparatus could be used to study nucleation, growth, recrystallization in impurity rich solutions \cite[e.g.][]{Peppin2009}, and protein interactions with ice surfaces \cite{Pertaya2007a, Pertaya2007b}.  In addition to other cryobiological settings \cite{Mohlmann2009}, crystal surface roughness and behavior far from equilibrium could also be investigated. 

The apparatus we have designed allows researchers a new level of flexibility when attempting to study complex polycrystalline materials.   Continuous monitoring and control allow for a range of potential experiments, very close to and far from equilibrium.  Most importantly it makes possible, a new and direct measurement of grain boundaries.   
 
\appendix*
\section{Temperature Model}

The thinness of the ice cell, creates an essentially two dimensional temperature field that can be calculated exactly, knowing the cooler tip temperature ($T_c$), heater temperature ($T_h$), and ice disk radius ($R_i$).  We model the system using a steady state, two dimensional, radial diffusion equation,
\begin{eqnarray}
\label{eq:diffQ} \frac{ \kappa _{(i,w)}}{r}\left( \frac{\partial}{\partial r} r \frac{\partial T}{\partial r}\right) = Q,\\
\label{eq:heatcon} k_w\frac{\partial T}{\partial r}| _{R_{i+}} = k_i\frac{\partial T}{\partial r}| _{R_{i-}},
\end{eqnarray}
where the interfacial boundary condition \eref{eq:heatcon} describes energy conservation at the solid/liquid interface.  The thermal diffusivity of the region of interest $\kappa_{(i,w)}=k_{(i,w)}/ \rho_{(i,w)} C_p^{(i,w)}$, where $k_i$ ($k_w$) is the thermal conductivity of ice (water), $\rho_i$ ($\rho_w$) and $C_p^i$ ($C_p^w$) are the density the heat capacity of ice (water).  An additional term, $Q$, is included in the diffusion equation \eref{eq:diffQ} to account for residual ambient heat flux from the laboratory.  The general solution to \eref{eq:diffQ} is found by integrating twice, $T(r)=A\ln r+B -\frac{Qr^2}{4\kappa}$.  Within the solid, $T=T_c$ at $r=R_c$ and  $T=T_m$ at $r=R_i$,
\begin{eqnarray}
\label{eq:ice1}T_c=A_i \ln R_c+ B_i- \frac{QR_c^2}{4\kappa_i},\\\
\label{eq:ice2}T_m=A_i \ln R_i+ B_i- \frac{QR_i^2}{4\kappa_i},
\end{eqnarray}
where $T_m$ is the melting temperature.  Here we assume that within the cooler tip radius, $R_C$, the temperature within the cell is isothermal.  Find $A_i$ and $B_i$ by subtracting \eref{eq:ice2} from \eref{eq:ice1} and substituting.  The result is a general expression for temperature within the ice,
\begin{multline}
\label{eq:icefull}T_i(r)=\left[ \frac{(T_c-T_m)}{\ln R_c/R_i}-\frac{Q (R_c^2-R_i^2)}{4\kappa_i \ln R_c/R_i} \right]
( \ln r- \ln R_c)\\
 +  \frac{Q(r^2-R_c^2)}{4\kappa_i} +T_c.
\end{multline}
Normalizing the chamber's radius to one leads to a similar system of equations within the liquid, where the boundary conditions are,  $T=T_m$ at $r=R_i$  and $T=T_h$ at $r=1$,
\begin{eqnarray}
\label{eq:liq1}T_m=A_w \ln R_i+ B_w- \frac{QR_i^2}{4\kappa_w},\\
\label{eq:liq2}T_h= B_w- \frac{Q}{4\kappa_i}.
\end{eqnarray}
Whence we find an expression for the temperature field within the water,
\begin{equation}
\label{eq:liqfull}T_w(r)=\left[ \frac{T_m-T_h}{\ln R_i}+\frac{Q(1-R_i^2)}{4\kappa_w \ln R_i} \right]  \ln r +  \frac{Q(r^2-1)}{4\kappa_w}+T_h.
\end{equation}
With the addition of the unknown radiative term, $Q$, the matching condition \eref{eq:heatcon} at the interface yields a transcendental relationship between the ice disk radius $R_i$ and $Q$,   
\begin{multline} \label{eq:RiQ}
(k_i\kappa_w-\kappa_ik_w)QR_i^2 \ln R_i \ln (R_c/R_i)\\
=2\kappa_i\kappa_wk_w(T_m-T_h) \ln(R_c/R_i)+2k_i\kappa_i\kappa_w(T_m-T_c)\ln R_i\\
+\kappa_ik_wQ(1-R_i^2) \ln (R_c/R_i)/2+k_i\kappa_wQ(R_c^2-R_i^2) \ln R_i /2.
\end{multline}
The flux $Q$ thus acts as a correction factor that we determine using observed ice disk radii, captured from time lapse photography while simultaneously measuring cooler tip and heater ring temperatures.  Ice disk radii and temperatures were recorded for multiple ice disk samples at multiple times and an average flux, $\bar{Q}= -7.957 \times 10^{-6} K/s$, was found with a standard deviation of, $\sigma_{\bar{Q}}=3.502 \times 10^{-8} K/s$. 

Utilizing $\bar{Q}$ and the experimental parameters we recursively solve the transcendental equation \eref{eq:RiQ} for $R_i$ and can use this to solve \eref{eq:icefull} at a given radius, $r$.  The radius of primary interest is the radius at which the beam measurement is being made, which is recorded from the vertical translation stage position.  In this manner temperature at the measurement position can be calculated, at the $95\%$ confidence levels, to within $0.02 K$. 

\begin{acknowledgments}
The authors are pleased to acknowledge conversations with J.G. Dash that led to a refined version of the ice growth cell component of this apparatus and thank D. DeMille for the use of the power meter for calibration.  We thank the Leonard X. Bosack and Bette M. Kruger foundation, the US National Science Foundation (No. OPP0440841), the Department of Energy (No. DE-FG02-05ER15741), the Helmholtz Gemeinschaft Alliance, `Planetary evolution and Life', and Yale University for generous support of this research.  The work of L.A. Wilen was supported by the US National Science foundation (OPP-0135989 and ANT-0439805).
\end{acknowledgments}


\end{document}